\documentclass[12pt]{article}
\usepackage {amssymb}
\usepackage {amsmath}
\usepackage[dvips]{epsfig}

\begin{document}

\begin{center}

\Large\bf{Change in the microscopic diffusion mechanisms of boron
implanted into silicon with increase in the annealing temperature}
\\[2ex]

\normalsize
\end{center}

\begin{center}
\textbf{O. I. Velichko and A. A. Hundorina}
\end{center}

Department of Physics, Belarusian State University of Informatics
and Radioelectronics, 6, P.~Brovka Street, Minsk, 220013 Belarus

{\it E-mail addresses: velichkomail@gmail.com} (Oleg Velichko);

{\it lena\_gundorina@mail.ru} (Alena Hundorina)

\begin{abstract}

A two stream model of boron diffusion in silicon has been
developed. The model is intended for simulation of transient
enhanced diffusion including redistribution of ion-implanted boron
during low temperature annealing. The following mechanisms of
boron diffusion were proposed, namely: the mechanism of a
long-range migration of nonequilibrium boron interstitials and the
mechanism due to the formation, migration, and dissolution of the
``impurity atom —-- silicon self-interstitial'' pairs. Based on
the model, simulation of the redistribution of boron implanted
into silicon substrates for annealing temperatures of 800 and 900
Celsius degrees was carried out. The calculated boron
concentration profiles agree well with the experimental data. It
was shown that for a temperature of 800 Celsius degrees the
transport of impurity atoms occurred due to the long-range
migration of nonequilibrium boron interstitials generated during
cluster transformation or dissolution. On the other hand, it was
found that at a temperature of 900 Celsius degrees the pair
diffusion mechanism played a main role in the significant
transient enhanced diffusion. A number of parameters describing
the transport of nonequilibrium boron interstitials and transient
enhanced diffusion of substitutionally dissolved boron atoms were
determined. For example, it was found that at a temperature of 900
Celsius degrees the time-average enhancement of boron diffusion
was approximately equal to 44 times. The results obtained are
important for the development of methods of transient enhanced
diffusion suppression keeping in mind the scaling of the
dimensions of silicon integrated microcircuits.

\end{abstract}



\section{Introduction}

At present the high-dose low-energy ion implantation is widely
used for manufacturing silicon  microcircuits with
ultra-large-scale integration (ULSI). During annealing of
ion-implanted layers, transient enhanced diffusion (TED) of dopant
atoms occurs. The TED is one of the major factors that limits the
device scaling and parameter improvement, especially for the
active regions with the {\it p}-type of conductivity formed by B
implanted into Si with doses providing dopant concentration above
the solubility limit for the annealing temperature. To suppress
the TED of boron atoms, various methods are used in the technology
of ULSI, including dopant implantation into the preamorphized
layer \cite{Hamilton_07,Yeong_08}.

\begin{figure}[ht]
\centering {\includegraphics{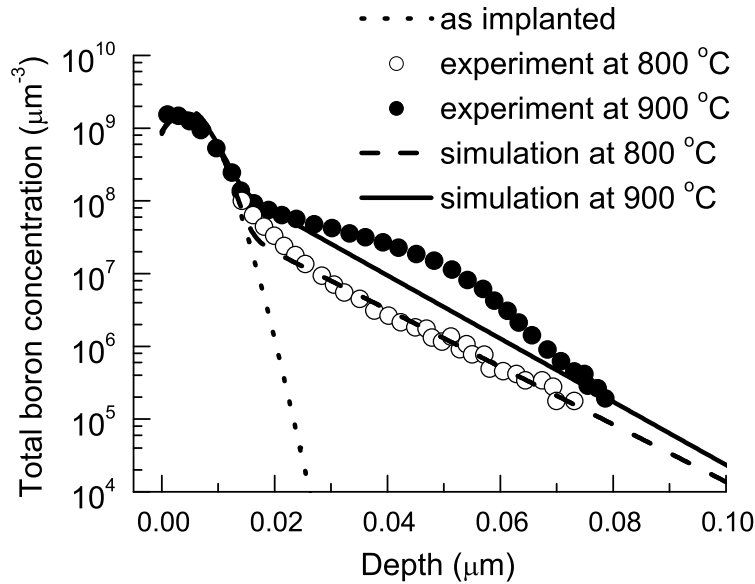}}

\caption{Total boron concentration profiles calculated for
temperatures of 800 and 900 $^{\circ}$C on the basis of the
long-range migration of boron interstitials. The experimental data
(open and black circles) are taken from Yeong {\it et al.}
\cite{Yeong_08}. \label{fig:Bint}}
\end{figure}

In Fig.~\ref{fig:Bint} the boron concentration profiles after
annealing for 60 s at temperatures of 800 and 900 $^{\circ}$C are
presented. The experimental boron distributions measured by
secondary ion mass spectroscopy (SIMS) are taken from Yeong {\it
et al.} \cite{Yeong_08}. In \cite{Yeong_08} Czochralski grown
(100) {\it n}-type silicon wafers were preamorphized by germanium
(Ge) ion implantation at 15 keV to a dose of 1.5$\times$10$^{15}$
cm$^{-2}$. Thereafter, boron implantation was conducted at 1 keV
with the same dose. The thermal annealing was carried out under
N$_{2}$ ambient, with ramp-up and ramp-down rates being equal to
60 $^{\circ}$C/s and 45 $^{\circ}$C/s, respectively.

It is seen from Fig.~\ref{fig:Bint} that significant diffusion
redistribution of the concentration profiles of ion-implanted
boron occurs during rapid thermal annealing when part of impurity
atoms is transferred into the undoped region located beneath the
surface. Based on the experimental results of \cite{Yeong_08}, it
is possible to formulate the following features of ion-implanted
boron redistribution: (i). The main amount of boron atoms located
in the region of high impurity concentration is immobile. Only
negligible diffusion near the surface occurs that does not change
the position of the peak of impurity concentration formed by ion
implantation. (ii). Significant redistribution of boron is
observed only in the region of concentrations below 10$^{8}$
$\mu$m$^{-3}$. (iii). Boron redistribution in the low
concentration region differs at temperatures of 800 and 900
$^{\circ}$C, namely: a) at a temperature of 800 $^{\circ}$C the
boron concentration profile after annealing has an extended
``tail'', and the shape of this "tail" is a straight line if the
axis of concentration is logarithmic; b) after annealing at a
temperature of 900 $^{\circ}$C the profile of boron distribution
in the region of concentrations below 10$^{8}$ $\mu$m$^{-3}$
rounded upward. It may be presumed that the different shape of
boron concentration profiles for the temperatures of 800 and 900
$^{\circ}$C is due to the change in the microscopic mechanism of
boron diffusion on increase in the annealing temperature.

Thus, the analysis of the experimental data of \cite{Yeong_08}
allows us to formulate the goal of the present work: to
investigate the microscopic mechanisms of boron transport in the
temperature range 800 -- 900 $^{\circ}$C when the change in the
microscopic mechanisms occurs.

\section{Microscopic mechanisms of boron diffusion}

It was supposed in \cite{Velichko_83} that the formation of
extended ``tails'' in the region of low impurity concentration
during short low temperature thermal treatments of ion-implanted
layers occurred due to migration of nonequilibrium impurity
interstitials. Analysis of the analytical solutions
\cite{Velichko_07,Velichko_10} obtained for the case of continuous
generation of nonequilibrium impurity interstitials within an
implanted layer shows that there occurs the formation of a
``tail'' in the low-concentration region of impurity atoms during
annealing. Moreover, the impurity concentration profile in this
low concentration region represents a straight line if the axis of
concentration is logarithmic. This conclusion is confirmed by the
calculations made in \cite{Velichko_10a}.

It maybe assumed that during the recrystallization of the
amorphous layer at the initial stage of annealing the boron atoms
occupy the substitutional position. These atoms are immobile
within the implanted layer being displaced at the initial stage of
annealing only for a short distance at a temperature near or below
800 $^{\circ}$C to form clusters which incorporate boron atoms and
self-interstitials. During the subsequent annealing,
transformation and dissolution of these clusters occur. It is
supposed that during these processes part of dopant atoms becomes
interstitial and the other boron atoms return again to the
substitutional position \cite{Velichko_88}. These substitutional
atoms are immobile as before because the temperature is too low
for the formation and migration of pairs, including point defect
and dopant atom, whereas boron interstitials can migrate to the
surface and into the bulk of the semiconductor. Migrating into the
bulk, these boron interstitials form an extended ``tail'' in the
low-concentration region which is characterized by a straight line
if the logarithmic scale is used for impurity concentration.

At a temperature near 900 $^{\circ}$C or above the
substitutionally dissolved boron atoms become also mobile. Due to
the high concentration of nonequilibrium self-interstitials, a
significant transient enhanced diffusion occurs which provides the
formation of a ``tail'' rounded upward. It is supposed that
diffusion of substitutionally dissolved boron occurs due to the
formation, migration, and dissolution of the ``impurity atom
--- silicon self-interstitial'' pairs \cite{Velichko_88,Velichko_84}.
Note, that these pairs are in local equilibrium with the
substitutionally dissolved boron atoms and nonequilibrium
self-interstitials. We also suppose that on this stage of
annealing the vacancy concentration is negligible, and one can
neglect the diffusion flux occurring due to the boron pairing with
vacancies.

\section{Model of boron diffusion}

The analysis of the microscopic mechanisms allows us to propose
the following model of ion-implanted boron diffusion. It is seen
from Fig.~\ref{fig:Bint} that irrespective of the temperature, the
main part of boron atoms in the region of high dopant
concentration is immobile and forms clusters with silicon
self-interstitials \cite{Pelaz_99}. During transformation or
dissolution of these clusters at a temperature near 900
$^{\circ}$C or above one fraction of boron atoms becomes
substitutional and can diffuse by the mechanism of the formation,
migrations and dissolution of the ``impurity atom —-- silicon
self-interstitial'' pairs, whereas the other fraction of
previously clustered boron occupies the interstitial position and
can migrate due to changing the interstitial sites in silicon
lattice. Taking into account the transformation or dissolution of
boron clusters, one can use the following systems of equations to
describe these processes of boron diffusion:

\textbf{1. Expression for the total concentration of impurity
atoms $C^{T}$}:

\begin{equation}\label{Total}
 C^{T} = C + C^{AI}+ C^{AC} + C^{AD} {\rm ;}
\end{equation}

\textbf{2. Conservation law for the impurity atoms incorporated
into clusters} \cite{Velichko_88}:

\begin{equation}\label{Cluster}
\frac{{\partial \,C^{AC}}}{{\partial \,t}} = S^{ACS} + S^{ACI} -
G^{ACS} - G^{ACI} {\rm ;}
\end{equation}

\textbf{3. Conservation law for the impurity atoms bound to
extended defects}  \cite{Velichko_88}:

\begin{equation}\label{Extended}
\frac{{\partial \,C^{AD}}}{{\partial \,t}} = S^{ADS} + S^{ADI} -
G^{ADS} - G^{ADI} {\rm ;}
\end{equation}

\textbf{4. Conservation law for all impurity atoms:}

\begin{equation}
\label{ConservLow} {\frac{{\partial C^{T}(x,t)}}{{\partial t}}} =
\displaystyle{\frac{{\partial \,C}}{{\partial \,t}}}
 + S^{AS} + S^{AI} - G^{AS} - G^{AI} {\rm ;}
\end{equation}

\textbf{5. Equation describing the pair diffusion mechanisms}
\cite{Velichko_88}:
\begin{equation}
\label{PairDifEq}
\begin{array}{l}
 \displaystyle{\frac{{\partial \,C}}{{\partial \,t}}} = \displaystyle{{\frac{{\partial} }{{\partial x}}}}
 {\left[ {D^{E}\;{\frac{{\partial
\left( {\tilde {C}^{V\times} C} \right)}}{{\partial x}}} +
{\frac{{D^{E}\tilde {C}^{V\times} C}}{{\chi}
}}\;\;{\frac{{\partial \chi} }{{\partial x}}}} \right]} \\
 \\ \quad \quad +  \displaystyle{{\frac{{\partial} }{{\partial x}}}}{\left[
{D^{F}\;{\frac{{\partial \left( {\tilde {C}^{I\times} C}
\right)}}{{\partial x}}} + {\frac{{D^{F}\tilde {C}^{I\times}
C}}{{\chi }}}\;\;{\frac{{\partial \chi} }{{\partial x}}}} \right]}
+{\frac{{C^{AI}(x,t)}}{{\tau ^{AI}}}} - S^{AS} + G^{AS} {\rm \rm ;}\\
 \end{array}
\end{equation}

\textbf{6. Equation describing the long-range migration of
nonequilibrium boron interstitials}
\cite{Velichko_10a,Velichko_88}:

\begin{equation}
\label{IntDifEq} {\frac{{\partial ^{\, 2}C^{AI}}}{{\partial
x^{2}}}} - {\frac{{C^{AI}}}{{l_{AI}^{2}} }} -
{\frac{{S^{AI}_{\tau}}}{{l_{AI}^{2}} }} +
{\frac{{G^{AI}_{\tau}}}{{l_{AI}^{2}} }} = 0 {\rm \rm ,}
\end{equation}

\noindent where

\begin{equation}\label{InterstitionalRates}
 S^{AI} = S^{ACI} + S^{ADI} \quad {\rm ,} \qquad G^{AI} = G^{ACI} + G^{ADI} \quad {\rm ,}
\end{equation}

\begin{equation}\label{SubstitutionalRates}
\quad  S^{AS} = S^{ACS} + S^{ADS}{\rm ,} \qquad  G^{AS} = G^{ACS}
+ G^{ADS}\quad {\rm ,}
\end{equation}

\begin{equation} \label{MigrationLenght}
l_{AI} = \sqrt {d^{AI}\tau ^{AI}} ,
\end{equation}

\begin{equation}
\label{NormRates} S^{AI}_{\tau}=S^{AI}\tau^{AI}\quad {\rm ,}
\qquad G^{AI}_{\tau} = G^{AI}\tau ^{AI} {\rm .}
\end{equation}

Here $C$ and $C^{AI}$ are the concentrations of substitutionally
dissolved impurity atoms and nonequilibrium dopant interstitials,
respectively; $C^{AC}$ and $C^{AD}$ are the concentrations of
impurity atoms incorporated into clusters and bound to the
extended defects, respectively; $S^{ACS}$ and $S^{ACI}$ are
respectively the rates of absorption of substitutionally dissolved
impurity atoms and impurity interstitials due to the cluster
formation; $S^{ADS}$ and $S^{ADI}$ are respectively the rates of
absorption of substitutionally dissolved impurity atoms and
impurity interstitials by extended defects; $G^{ACS}$ and
$G^{ACI}$ are respectively the rates of generation of separate
substitutionally dissolved impurity atoms and impurity
interstitials during cluster transformation or dissolution;
$G^{ADS}$ and $G^{ADI}$ are respectively the rates of generation
of separate substitutionally dissolved impurity atoms and impurity
interstitials during extended defect annealing; $\tilde
{C}^{V\times}$ and $\tilde {C}^{I\times }$ are the concentrations
of vacancies and self-interstitials in the neutral charge state
normalized to the equilibrium concentrations $C_{eq}^{V\times} $
and $C_{eq}^{I\times}$, respectively; $D^{E}\left( {\chi} \right)$
is the effective diffusivity of impurity atoms due to the
vacancy---impurity pair mechanism; $D^{F}\left( {\chi} \right)$ is
the effective diffusivity of impurity atoms due to migration of
the ``impurity atom---self-interstitials'' pairs; $\chi $  is the
concentration of charge carriers normalized to the intrinsic
carrier concentration $n_{i} $; $C^{B}$ is the concentration of
impurity with the opposite-type conductivity; $d^{AI}$ is the
diffusivity of nonequilibrium impurity interstitials; $\tau ^{AI}$
is the average lifetime of impurity interstitials mediated by
recombination with vacancies and kickout of silicon atoms from the
lattice sites. We note that the concentration dependencies
$D^{E}\left( {\chi} \right)$ and $D^{F}\left( {\chi} \right)$ are
described in \cite{Velichko_08}.

\section{Results of numerical calculations}

In this section we use the developed model of boron diffusion for
simulation of the experimental data of Yeong {\it et al.}
\cite{Yeong_08}. It is worth to note that the experimental boron
profiles measured in \cite{Yeong_08} for annealing temperatures of
800 and 900 $^{\circ}$C are characterized, in contrast to the work
of Hamilton {\it et al.} \cite{Hamilton_07}, by the absence of a
local peak of boron concentration in the region of EOR defects.
Therefore, we neglect all the processes mediated by the extended
defects. We also suppose that after the initial stage of annealing
only transformation or dissolution of the boron clusters occurs,
and the absorption of boron atoms by the clusters is negligible.

To describe the spatial distribution of impurity atoms after solid
phase recrystallization $C_{0} (x)$and the spatial distribution of
the generation rate for boron interstitials $G^{AI}(x,t)$, the
Pearson-IV distribution $f^{P}(x,R_{p},\Delta R_{p},S_{k})$
\cite{Burenkov_85} is used:

\begin{equation} \label{CTIni}
C_{0} (x) = C_{m}f^{P}(x,R_{p},\Delta R_{p},S_{k}) {\rm ,}
\end{equation}

\begin{equation} \label{GAI}
G^{AI}(x,t) = g_{m}f^{P}(x,R_{p},\Delta R_{p},S_{k}) {\rm ,}
\end{equation}

\noindent where

\begin{equation} \label{Cmax}
C_{m} = {\frac{{Q}}{{\sqrt {2\pi} \Delta R_{p}} }}\times 10^{ -
8}{\rm .}
\end{equation}

Here $C_{m}$ is the maximal concentration of boron atoms after
implantation; $g_{m}$ is the maximal value of the generation rate
of boron interstitials per unit volume; $Q$ $[cm^{-2}]$ is the
dose of ion implantation; $R_{p}$ and $\Delta R_{p}$ [$\mu$m] are
the average projective range of boron ions and straggling of the
projective range, respectively; $S_{k}$ is the skewness of the
impurity profile.

In Fig.~\ref{fig:Bint} the boron concentration profiles calculated
for the diffusion occurring only due to the long--range migration
of nonequilibrium boron interstitials are shown. As can be seen
from Fig.~\ref{fig:Bint}, the calculated  boron profile is in good
agreement with the experimental data for an annealing temperature
of 800 $^{\circ}$C. The following values of the model parameters
were used to provide the best fit of the calculated boron
concentration profile to the experimental one:

\textbf{Parameters prescribing the initial distribution of
implanted boron:} $Q$=1.5$\times$10$^{15}$ cm$^{-2}$;  $R_{p}$ =
0.0044 $\mu$m; $\Delta R_{p}$ = 0.0036 $\mu$m: $S_{k}$ = 0.24.

\textbf{Parameters specifying  the process of interstitial
diffusion:} the duration of annealing  $\tau _{p}$=60 s; the
annealing temperature $T_{C}$=800\,$^{\circ}$C; the maximum value
of the generation rate of nonequilibrium impurity interstitials
$g_{m}$ = 2.4$\times$10$^{6}$ $\mu$m$^{-3}$\,s$^{-1}$; the average
migration length of boron interstitials $l^{AI}$ = 11 nm; the
effective escape velocity of interstitial impurity atoms
$\mathrm{v}^{S}_{eff}$= 0; the concentration of boron
interstitials on the right boundary $C_{B}^{AI}$=0; the position
of the right boundary $x_{B}$=0.5 $\mu$m. It is also supposed that
the average lifetime of nonequilibrium impurity interstitials
$\tau ^{AI}$ is significantly shorter than the duration of
annealing.

The boron concentration profile presented in Fig.~\ref{fig:Bint}
for a temperature of $T_{C}$=800\,$^{\circ}$C was calculated on
the assumption that approximately 6.9 \% of the implanted boron
atoms are being transferred to the transient interstitial
positions and then become immobile again occupying the
substitutional sites. Migration of these nonequilibrium
interstitial atoms results in the formation of an extended ``tail"
on the boron concentration profile.

A similar simulation was also performed for a temperature of
$T_{C}$=900 $^{\circ}$C. The following parameters specifying the
process of interstitial diffusion were used: $g_{m}$ =
9.0$\times$10$^{6}$ $\mu$m$^{-3}$\,s$^{-1}$; $l^{AI}$ = 10 nm. It
can be seen from Fig.~\ref{fig:Bint} that the calculated boron
profile disagrees with the experimental one. The quantitative
difference of the shape of the experimental boron concentration
profile points to the fact that boron diffusion at a temperature
of $T_{C}$=900 $^{\circ}$C cannot be attributed to the mechanism
of the log--range migration of nonequilibrium impurity
interstitials. Therefore, we curried out a new simulation taking
into account two independent fluxes of boron atoms.

\begin{figure}[ht]
\centering {\includegraphics{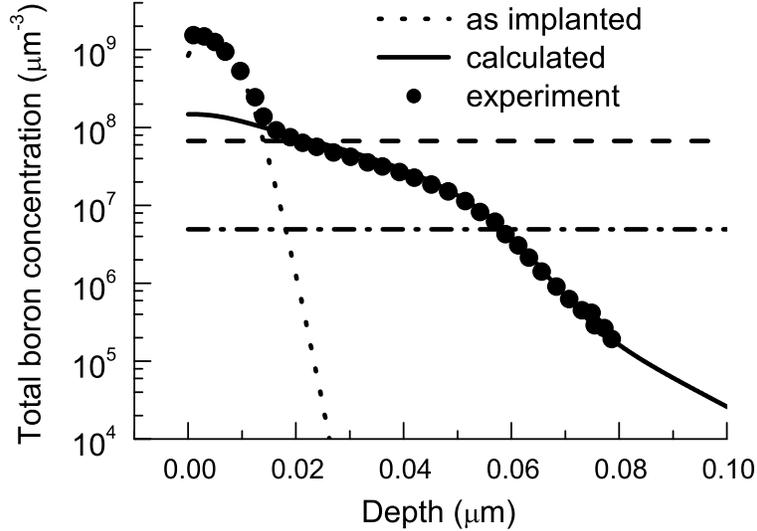}}

\caption{Substitutionally dissolved boron concentration profile
calculated for a temperatures of 900 $^{\circ}$C on the basis of
the model considering two mechanisms of boron diffusion. The
experimental data for total boron concentration (black circles)
are taken from Yeong {\it et al.} \cite{Yeong_08}. Dashed curve
--- boron solubility limit in silicon. The dashed-dotted curve ---
concentration of intrinsic carriers.} \label{fig:Btwo}
\end{figure}

Figure \ref{fig:Btwo} presents the profiles of the total boron
concentration and concentration of substitutionally dissolved
boron atoms after annealing at 900 \,$^{\circ}$C calculated within
the framework of the full two stream diffusion model. It is
supposed that the pair diffusion mechanism play an important role
at a temperature of 900 \,$^{\circ}$C in addition to the
long-range migration of boron interstitials.  The value of boron
diffusivity obtained from the best fit to the experimental data is
equal to 2.7$\times$10$^{-6}$ $\mu$m$^{2}$/s that approximately 46
times exceeds the thermal equilibrium value of diffusivity equal
to 6.11$\times$10$^{-8}$ $\mu$m$^{2}$/s. It means that a
significant transient enhanced diffusion occurs. The extracted
value of the empirical coefficient $\beta^{F}_{1}$ which describes
the contribution of singly charged self-interstitials
\cite{Velichko_08} is equal to 0.27 a.u. The average migration
length of boron interstitials is equal to 12 nm.

The good agreement of the calculated boron concentration profile
with the experimental data in contrast to Fig.~\ref{fig:Bint}
allows one to make a conclusion that at 900 \,$^{\circ}$C the
basic mechanism of the boron transient enhanced diffusion is the
mechanism of formation, diffusion, and dissolution of the
``substitutionally dissolved boron atom
--- silicon self-interstitial'' pairs, whereas at a temperature of 800
\,$^{\circ}$C the impurity transport is due only to the long-range
migration of nonequilibrium boron interstitials.

\newpage

\section{Conclusions}

The model of boron diffusion in silicon has been developed. The
model is intended for simulation of transient enhanced diffusion
including the redistribution of ion-implanted boron during low
temperature annealing. Two possible mechanisms of boron diffusion
are taken into account, namely: the mechanism of the long-range
migration of nonequilibrium boron interstitials and the mechanism
of the formation, migration, and dissolution of the ``impurity
atom —-- silicon self-interstitial'' pairs. Based on the model
simulation of the redistribution of boron implanted in silicon
substrates with a low energy and a high dose was carried out. The
case of rapid thermal annealing (60 s) was investigated for two
characteristic temperatures of 800 and 900 $^{\circ}$C. The
calculated boron concentration profiles agree well with the
experimental data of \cite{Yeong_08}. It is shown that for a
temperature of 800 $^{\circ}$C the boron atoms substitutionally
dissolved in the silicon lattice are immobile and the transport of
impurity atoms occurred due to the long-range migration of
nonequilibrium boron interstitials formed during cluster
transformation or dissolution. On the other hand, it is shown that
at an annealing temperature of 900 $^{\circ}$C the mechanism of
formation, diffusion, and dissolution of the ``substitutionally
dissolved boron atom
--- silicon self-interstitial'' pairs plays the main role in the
significant transient enhanced diffusion. A number of parameters
describing the transport of nonequilibrium boron interstitials and
transient enhanced diffusion of substitutionally dissolved boron
atoms have been determined. For example, it is found that the
average migration length of nonequilibrium boron interstitials is
equal to 11 nm at a temperature of 800 $^{\circ}$C. At a
temperature of 900 $^{\circ}$C the time-average boron diffusion
enhances approximately 44 times. The results obtained are
important for the development of the methods of transient enhanced
diffusion suppression keeping in mind the scaling of the
dimensions of silicon integrated microcircuits.

\section{Acknowledgement}
This work has been partially  supported by the Belarusian
Republican Foundation for Fundamental Research under Grant No.
F09-050.

\end{document}